\input Tex-document.sty
\gdef\preliminary{F}
\catcode`\{=1
\if T\preliminary
\openout\refs=references.aux
\write\refs{\string\relax}
\def\cite#1{%
\write\refs{\string\citation\string{#1\string}}
#1}
\def\endref{\write\refs{\string\bibstyle\string{china\string}}
\write\refs{\string\bibdata\string{refs\string}}}
\else
\expandafter\xdef\csname refEH2000\endcsname{1}\relax
\expandafter\xdef\csname refEH2001\endcsname{2}\relax
\expandafter\xdef\csname refEPR1999a\endcsname{3}\relax
\expandafter\xdef\csname refEPR1999b\endcsname{4}\relax
\expandafter\xdef\csname refECM1990\endcsname{5}\relax
\expandafter\xdef\csname refGC1995\endcsname{6}\relax
\expandafter\xdef\csname refH:2001\endcsname{7}\relax
\expandafter\xdef\csname refH2002\endcsname{8}\relax
\expandafter\xdef\csname refJP1998\endcsname{9}\relax
\expandafter\xdef\csname refK1998\endcsname{10}\relax
\expandafter\xdef\csname refLS1999\endcsname{11}\relax
\expandafter\xdef\csname refRT2000\endcsname{12}\relax
\expandafter\xdef\csname refRT2001\endcsname{13}\relax
\expandafter\xdef\csname refZabey2001\endcsname{14}\relax
\def\cite#1{\csname ref#1\endcsname}
\def\bibitem#1{\item{[\csname ref#1\endcsname]}}
\def\newblock{}
\def\endref{}
\def\begin#1#2{}
\fi
\let\rho=\varrho
\def\L{{\rm L}}
\def\R{{\rm R}}
\def\S{{\rm S}}

\def\hidekappa#1{}
\def\d{{\rm d}}
\def\dt{{\rm d}t}
\def\smm{\sum_{m=1}^M}
\def\Lm{{\L,m}}
\def\Rm{{\R,m}}
\def\frac#1#2{{#1\over #2}}
\def\eff{{\rm eff}}
\def\T{{\rm T}}

\def\SS{{\cal S}}

\def\dspq{\d s\,\d q\,\d p}

\def\Q(#1,#2){#1^{(#2)}}
\def\QP(#1,#2){\tilde #1^{(#2)}}
\def\Qp(#1){\tilde #1}
\def\LIKEREMARK#1{\smallskip\noindent{\bf #1}. }
\def\REMARKNR#1{
\NEWDEF{c}{#1}{Remark \actualnumber.\number\CLAIMcount}
\smallskip\noindent{\bf Remark \actualnumber.\number\CLAIMcount.~}%
\global\advance\CLAIMcount by 1}
\let\truett=\tt
\fontdimen3\tentt=2pt\fontdimen4\tentt=2pt
\def\tt{\hfill\break\null\kern -3true cm\truett
\#\#\#\#\#\#\#\#\#\#\#\#\#\#  }
\let\epsilon=\varepsilon

\def\CC{{\cal C}}

\def\SS{{\cal S}}

\def\em{\it}

\def\dx{{\rm d}x}

\def\dt{{\rm d}t}

\def\real{{\bf R}}
\def\CC{{\cal C}}
\def\S{{\rm S}}
\def\SS{{\cal S}}
\def\frac#1#2{{#1\over #2}}
\def\HH{{\cal H}}
\def\dx{{\rm d}x}

\def\d{{\rm d}}
\def\LL{{\cal L}}
\def\e{e}

\def\smm{\sum_{m=1}^M}
\def\im{{i,m}}
\def\Lm{{\L,m}}
\def\Rm{{\R,m}}
\def\smm{}
\def\im{{i}}
\def\Lm{{\L}}
\def\Rm{{\R}}
\def\const{{\rm const.}}
\def\HALF{{\textstyle{1\over 2}}}
\let\epsilon=\varepsilon
\let\kappa=\varkappa
\let\theta=\vartheta
\def\hidekappa#1{}
\def\norm{\vert\kern-0.1em\vert\kern-0.1em\vert}
\def\bnorm{\big\vert\kern-0.1em\big\vert\kern-0.1em\big\vert}
\newcount\EQNcount \EQNcount=1
\newcount\CLAIMcount \CLAIMcount=1
\newcount\SECTIONcount \SECTIONcount=1
\newcount\SUBSECTIONcount \SUBSECTIONcount=1

\def\ifff#1#2#3{\ifundefined{#1#2}%
\expandafter\xdef\csname #1#2\endcsname{#3}\else%
\immediate\write16{!!!!!doubly defined #1,#2}\fi}
\def\NEWDEF#1#2#3{\ifff{#1}{#2}{#3}}
\def\actualnumber{\number\SECTIONcount}
\def\EQ#1{\eqno\tageck{#1}}
\def\tageck#1{\lmargin{#1}({\rm \actualnumber}.\number\EQNcount)
 \NEWDEF{e}{#1}{(\actualnumber.\number\EQNcount)}
\global\advance\EQNcount by 1
}

\def\CLAIM#1#2#3\par{
\smallskip\noindent {\lmargin{#2}\bf #1\
\actualnumber.\number\CLAIMcount.} {\sl #3}\par \NEWDEF{c}{#2}{#1\
\actualnumber.\number\CLAIMcount} \global\advance\CLAIMcount by 1
\ifdim\lastskip<\medskipamount
\removelastskip\penalty55\medskip\fi}
\def\CLAIMNONR #1#2#3\par{
\smallskip\noindent {\lmargin{#2}\bf #1.} {\sl #3}\par
\NEWDEF{c}{#2}{#1} \global\advance\CLAIMcount by 1
\ifdim\lastskip<\medskipamount
\removelastskip\penalty55\medskip\fi}
\fontdimen16\tensy=2.7pt
\fontdimen13\tensy=4.3pt
\fontdimen17\tensy=2.7pt
\fontdimen14\tensy=4.3pt
\fontdimen18\tensy=4.3pt
\def\SECTION#1{    \global\advance\SECTIONcount by 1
{\head{\actualnumber.\ #1}}
    \EQNcount=1
    \CLAIMcount=1
    \SUBSECTIONcount=1
    \nobreak\sectionskip\noindent}
\def\SECTIONNONR#1{\vskip0pt plus.3\vsize\penalty-75
    \vskip0pt plus -.3\vsize
    \global\advance\SECTIONcount by 1
    \beforesectionskip\noindent
{\sectionsize\sectiontype  #1}
     \EQNcount=1
     \CLAIMcount=1
     \SUBSECTIONcount=1
     \nobreak\sectionskip\noindent}
\def\SUBSECTION#1{\vskip0pt plus.2\vsize\penalty-75%
    \vskip0pt plus -.2\vsize%
    \beforesectionskip\noindent%
{\subsectionsize\subsectiontype \actualnumber.\number\SUBSECTIONcount.\ #1}
    \global\advance\SUBSECTIONcount by 1
    \nobreak\sectionskip\noindent}
\def\SUBSECTION#1{\global\advance\SUBSECTIONcount by 1
\head{\subsectionsize\subsectiontype \actualnumber.\number\SUBSECTIONcount.\ #1}
}
\def\SUBSECTIONNONR#1\par{\vskip0pt plus.2\vsize\penalty-75
    \vskip0pt plus -.2\vsize
\beforesectionskip\noindent
{\subsectionsize\subsectiontype #1}
    \nobreak\sectionskip\noindent\noindent}
\def\CLAIM#1#2#3\par{
\smallskip\noindent {\lmargin{#2}\bf #1\
\actualnumber.\number\CLAIMcount.} {\sl #3}\par \NEWDEF{c}{#2}{#1\
\actualnumber.\number\CLAIMcount} \global\advance\CLAIMcount by 1
\ifdim\lastskip<\medskipamount
\removelastskip\penalty55\medskip\fi}
\def\CLAIMNONR #1#2#3\par{
\smallskip\noindent {\lmargin{#2}\bf #1.} {\sl #3}\par
\NEWDEF{c}{#2}{#1} \global\advance\CLAIMcount by 1
\ifdim\lastskip<\medskipamount
\removelastskip\penalty55\medskip\fi}
\def\ifundefined#1{\expandafter\ifx\csname#1\endcsname\relax}
\def\equ#1{\ifundefined{e#1}$\spadesuit$#1\else\csname e#1\endcsname\fi}
\def\clm#1{\ifundefined{c#1}$\spadesuit$#1\else\csname c#1\endcsname\fi}
\def\sec#1{\ifundefined{s#1}$\spadesuit$#1
\else Section \csname s#1\endcsname\fi}
\def\lab#1#2{\ifundefined{#1#2}$\spadesuit$#2\else\csname #1#2\endcsname\fi}
\def\fig#1{\ifundefined{fig#1}$\spadesuit$#1\else\csname fig#1\endcsname\fi}
\def\lmargin#1{}
\def\L{{\rm L}}
\def\R{{\rm R}}
\newdimen\papwidth
\newdimen\papheight
\newskip\beforesectionskipamount  
\newskip\sectionskipamount 
\def\sectionskip{\vskip\sectionskipamount}
\def\beforesectionskip{\vskip\beforesectionskipamount}

\def\sectionsize{\larger}
\def\sectiontype{\bf}
\def\subsectionsize{}
\def\subsectiontype{\bf}
\def\em{\sl}

\def\TTH0{T^t_{\HH_0}}


\def\firstpage{409}

\pageno=409

\def\shorttitle{Non-Equilibrium Steady States}
\def\shortauthor{J.-P. Eckmann}

\title{\centerline{Non-Equilibrium Steady States}}

\author{J.-P. Eckmann\footnote{\eightrm *}{\eightrm Section de
Math\'ematiques
et D\'epartement de Physique Th\'eorique,
Universit\'e de Gen\`eve, 1211 Geneva 4, Switzerland.
E-mail: eckmann@physics.unige.ch}}

\vskip 7mm

\centerline{\boldnormal Abstract}

\vskip 4.5mm

{\narrower \ninepoint \smallskip
The mathematical physics of mechanical systems in thermal equilibrium
is a well studied, and relatively easy, subject, because the Gibbs
distribution is in general an adequate guess for the equilibrium
state.

On the other hand, the mathematical physics of {\em
non-equilibrium} systems, such as that of a chain of masses
connected with springs to two (infinite) heat reservoirs is more
difficult, precisely because no such {\em a priori} guess exists.

Recent work has, however, revealed that under quite general
conditions, such states can not only be shown to exist, but are {\em
unique}, using the H\"ormander conditions and
controllability. Furthermore, interesting properties, such as
energy flux, exponentially fast convergence to the unique state, and
fluctuations of that state have been successfully studied.

Finally, the ideas used in these studies can be extended to certain
stochastic PDE's using Malliavin calculus to prove regularity of the process.

\vskip 4.5mm

\noindent{\bf 2000 Mathematics Subject Classification:} 82C22,
60H15.

\noindent{\bf Keywords and Phrases:} Non-equilibrium statistical mechanics, Stochastic differential equations.

}

\vskip 10mm

\head{1. The model and results}

I report here on work done, in different combinations, together with
{\em Martin Hairer, Luc Rey-Bellet, Claude-Alain Pillet, and Lawrence Thomas}.
In it, we considered the seemingly trivial problem of describing
the non-equilibrium statistical mechanics
of a finite-dimensional non-linear Hamiltonian system coupled to two
infinite heat reservoirs which are at {\em different} temperatures. By
this I mean that the stochastic forces of the two heat reservoirs
differ.
The difficulties in such a problem are related to the absence of an
easy {\it a priori} estimate for the state of the system.
We show
under certain conditions on the initial data that the system goes to a
{\em unique} non-equilibrium steady state and we describe rather
precisely some properties of this steady state. These are,

\item{$\bullet$}{appearance
of an energy flux (from the hot to the cold reservoir) whenever the reservoirs
are at different temperatures,}
\item{$\bullet$}{exponential stability of this state,}
\item{$\bullet$}{fluctuations around this state satisfy the
Cohen-Gallavotti fluctuation conjecture.}

I first review the construction of the model. Two
features need special attention: The modeling of the heat reservoirs
and their coupling to the chain, and the nature of the coupling among
the masses in the chain. I start with the latter:
It is a 1-dimensional chain of $n$ distinct $d$-dimensional anharmonic
oscillators with nearest neighbor coupling. The phase space
of the chain is therefore $\real^{2dn}$ and its
dynamics is described by a
$\CC^{\infty}$ Hamiltonian function of the form
$$
H_\S (p,q)\,=\,\sum_{j=1}^n \frac{p_j^2}{2} + \sum_{j=1}^n U_j^{(1)}(q_j)
+ \sum_{i=1}^{n-1} U^{(2)}_{i}(q_i-q_{i+1})\,\equiv\,\sum_{j=1}^n \frac{p_j^2}{2} +V(q)~,
\EQ{hamil}
$$
where $q=(q_1,\dots,q_n)$, $p=(p_1,\dots,p_n)$,
with $p_i$, $q_i \in \real^{d}$.
We will eventually couple the ends of the chain, {\it i.e.}, $q_1$ and
$q_n$,
to heat reservoirs. Clearly, for heat conduction to be possible at
all we must require that the $U_{i}^{(2)}$ are non-zero. But,
this is not enough and the interaction has to have a minimal
strength.
A sufficient condition for the main result
to hold is:
For some $m_2\ge m_1\ge2$, and all sufficiently large $|q|$, we require
$$
0\,<\,c_1\,\le\, {U_i^{(1)}(q)\over
(1+|q|)^{m_1}}\,\le\,c_1'~,\qquad
0\,<\,c_2\,\le\, {U_i^{(2)}(q)\over
(1+|q|)^{m_2}}\,\le\,c_2'~,
$$
and similar growth conditions on the first and second derivatives.
Finally, we require that
each of the ($d\times d$) matrices
$$
\nabla_{q_i}\nabla_{q_{i+1}}U^{(2)}_i(q_i-q_{i+1})~,\quad
i=1,\dots,n-1~,
\EQ{convex}
$$
is
non-degenerate (see [\cite{RT2000}] for the most general conditions).

\vskip2.7cm \includegraphics{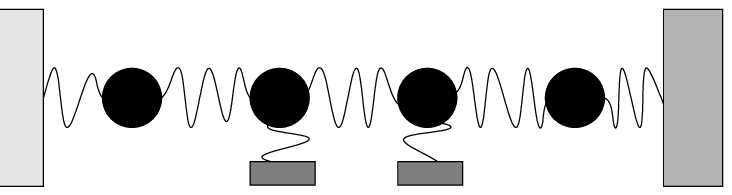}
\vskip-0.3cm

{\narrower \narrower \noindent{\bf Fig. 1}: Model of the chain
with the two reservoirs at its ends,

\hskip 4mm ``Hot'' at left, ``Cold'' at right.

}

\REMARKNR{111}$\,$ It seems that relaxing the condition \equ{convex} poses hard technical problems, although, from
a physical point of view, allowing the matrix to be degenerate on hyper-surfaces of codimension $\ge 1$ should
work. See [\cite{JP1998},\cite{RT2000}] for some possibilities.

\REMARKNR{2}$\,$ If $m_2<m_1$ it seems that the existence of a unique state is jeopardized by the potential
appearance of breathers. Indeed, too much energy can then be ``stored'' in $U^{(1)}$ without being sufficiently
``transported'' between the oscillators. A more detailed understanding of this problem would be welcome.

\REMARKNR{2a}$\,$ The nature of the steady state in the limit of an infinite chain ($n\to\infty $) is a difficult
open question.

As a model of a heat reservoir we consider the classical field theory associated
with the $d$-dimensional wave equation.
The field $\varphi$ and its conjugate momentum field $\pi$ are elements
of the real Hilbert space $\HH={\rm H}^1_{\real}(\real^{d}) \oplus
\L ^2_{\real}(\real^{d})$ which is the completion of $\CC_0^\infty
(\real^d)\oplus\CC_0^\infty
(\real^d)$
with respect to the norm defined by the scalar product:
$$
\left(\left(\matrix{\varphi\cr\pi\cr}\right),
\left(\matrix{\varphi\cr\pi\cr}\right)\right)_\HH
\,=\,\int {\dx}\,\left(|\nabla\varphi(x)|^2+|\pi(x)|^2\right)\,\equiv\,
2H_{\rm B}(\varphi,\pi)~,
\EQ{scalar}
$$
where $H_{\rm B}$ is the Hamiltonian of a bath
and the corresponding equation of motion is the ordinary wave equation
which we write in the form
$$
\left(\matrix{{\dot\varphi}(t)\cr{\dot\pi}(t)\cr}\right)\,=\,
\LL \left(\matrix{\varphi\cr\pi\cr}\right)\,\equiv\,\left(\matrix{0&1\cr\Delta&0\cr}\right)\left(\matrix{\varphi\cr\pi\cr}\right)~.
$$
Finally, we define the coupling between the chain and the heat reservoirs.
The reservoirs will be called ``$\L$'' and ``$\R$'', the left one
coupling to the coordinate $q_1$ and the right one coupling to the
other end of the chain ($q_n$).
Since we consider two heat reservoirs,
the phase space of the coupled system, for finite energy
configurations, is $\real^{2dn} \times {\cal{H}} \times {\cal{H}}$
and
its Hamiltonian will be chosen as
$$\eqalign{
H(p,q,\varphi_{\L},\pi_{\L},\varphi_{\R},\pi_{\R})\,&=\, H_\S (p,q) +
H_{\rm B}(\varphi_\L,\pi_\L) + H_{\rm B}(\varphi_\R,\pi_\R) \cr
&+
\hidekappa{\kappa_\L } q_1 \cdot \int {\dx}\, \rho_\L (x) \nabla \varphi_\L(x)
+ \hidekappa{\kappa_\R}  q_n \cdot \int {\dx} \,\rho_\R (x) \nabla
\varphi_\R(x)~.
}\EQ{coupham}
$$
Here, the \hidekappa{$\kappa_i$, $i\in\{\L,\R\}$
are real coupling constants and the }$\rho_i(x)\in\L^1(\real^d)$ are
charge densities which we assume for simplicity to be
spherically symmetric functions.
The choice of the Hamiltonian
\equ{coupham} is motivated by the dipole approximation of classical
electrodynamics. We use the shorthand
$$
\phi_i\equiv \left(\matrix{\varphi_i\cr\pi_i\cr}\right)~.
$$
We set
$\alpha_i=\left(\alpha_{i}^{(1)},\dots,\alpha_{i}^{(d)}\right)$,
$i\in\{\L,\R\}$, with
$$
{\widehat \alpha}_{i}^{(\nu)}(k)\equiv \left(\matrix{{-ik^{(\nu)}}
{\widehat \rho}_i(k)/k^2\cr 0\cr}\right)~.
$$
The ``hat'' means the Fourier transform
$
{\widehat f}(k)\,\equiv\,{(2\pi)^{-d/2}}\int {\dx}\,f(x)\e^{-ik\cdot
x}~.
$
With this notation the Hamiltonian is
$$
H(p,q,\phi_\L,\phi_\R)\,=\,
H_\S (p,q)+H_{\rm B}(\phi_\L)+H_{\rm B}(\phi_\R)+q_1 \cdot
(\phi_\L,\alpha _\L)_{\HH}+q_n \cdot
(\phi_\R,\alpha _\R)_{\HH}~,
$$
where $H_{{\rm B}}(\phi)=\HALF\|\phi\|_{\HH}^2$.
The equations of motions take the form
$$\eqalign{
{\dot q}_j (t) \,&=\, p_j (t)~, \qquad j =1,\dots,n~, \cr
{\dot p}_1(t) \,&=\, -\nabla_{q_1}V(q(t))-\hidekappa{\kappa_\L}
(\phi_\L(t),\alpha_\L )_{\HH}~, \cr
{\dot p}_j (t) \,&=\, -\nabla_{q_j }V(q(t))~, \qquad  j =2,\dots,n-1
{}~,\cr
{\dot p}_n(t) \,&=\, -\nabla_{q_n}V(q(t))-\hidekappa{\kappa_\R }
(\phi_\R(t),\alpha_\R)_{\HH} ~,\cr
{\dot \phi}_\L(t) \,&=\,\LL \left(\phi_\L(t) + \hidekappa{\kappa_\L}  \alpha_\L  \cdot
q_1(t)\right) ~,\cr
{\dot \phi}_\R(t) \,&=\, \LL \left(\phi_\R(t) +\hidekappa{ \kappa_\R}  \alpha_\R \cdot
q_n(t)\right)~.
}\EQ{eqmo1}
$$
The last two equations of \equ{eqmo1} are easily integrated and lead to
$$
\eqalign{
\phi_\L(t)\,&=\,\e^{\LL t}\phi_\L{(0)} +\hidekappa{ \kappa_\L}\int_0^t
\!\d s\,\LL e^{\LL(t-s)} \alpha_\L\cdot q_1(s)~, \cr
\phi_\R(t)\,&=\,\e^{\LL t}\phi_\R{(0)} +\hidekappa{ \kappa_\R}\int_0^t
\!\d s\,\LL e^{\LL(t-s)} \alpha_\R\cdot q_n(s)~, \cr
}
$$
where the $\phi_i{(0)}$, $i\in\{\L,\R\}$, are the initial conditions of the
heat reservoirs.

We next assume that the two reservoirs are in thermal equilibrium at
inverse temperatures
$\beta _\L$ and $\beta _\R$. By this I mean that the initial conditions
$\Phi(0) \equiv \{\phi_\L{(0)},\phi_\R{(0)}\}$
are random variables distributed according to a Gaussian measure with
mean zero and
covariance $\langle \phi_i(f)\phi_j(g)\rangle=\delta_{ij} (1/\beta_i)(f,
g)_{\HH}$.
If we assume that the coupling functions $\alpha_i^{(\nu)}$ are in
$\HH$ for $i \in \{\L,\R\}$ and $\nu \in \{1,\dots, d \}$,
then the $\xi_i(t) \equiv \phi_i{(0)}(e^{-\LL t}\alpha_i)$
are $d$-dimensional Gaussian random processes with mean zero and covariance
$$
\langle\, \xi_i(t) \xi_j(s) \rangle\,=\,\delta_{i,j}\frac{1}{\beta_i}
C_i(t-s)~,\quad i,j\in \{\L,\R\}~,
\EQ{cov1}
$$
where the $d\times d$ matrices $C_i(t-s)$ are
$$
C_{i}^{(\mu,\nu)}(t-s)
\,=\,\left(\alpha_{i}^{(\mu)},e^{\LL(t-s)}\alpha_{i}^{(\nu)}\right)_{\HH}
            \,=\,\frac{1}{d}\delta_{\mu,\nu}\int {\rm dk}\, |{\widehat
\rho_i}(k)|^2
                 \cos\bigl(|k|(t-s)\bigr)~.
$$

Finally, we impose a condition on the random force exerted by the heat
reservoirs
on the chain. We assume that
the covariances of the random processes $\xi_i(t)$ with
$i\in\{\L,\R\}$
satisfy
$$
C_i^{(\mu,\nu)}(t-s)\,=\,\delta_{\mu,\nu}\smm \lambda ^2_\im
e^{-\gamma_\im |t-s|}~,
\EQ{H3}
$$
with $\gamma_\im>0$
and $\lambda _\im>0$,
which can be achieved by a suitable choice of
the coupling functions $\rho_i(x)$, for example
$$
\widehat \rho
_i(k)\,=\,\const \prod_{m=1}^M{1\over (k^2 +\gamma_\im^2)^{1/2}}~,
\EQ{MM}
$$
where all the $\gamma_\im$ are distinct.
We continue with the case $M=1$ for simplicity.

Using \equ{H3}
and enlarging the phase space with auxiliary fields $r_i$,
one eliminates the memory terms (both deterministic and random) of
the equations of motion and rewrites them
as a system of {\em Markovian} stochastic differential equations:
$$\eqalign{
\d q_j (t) \,&=\, p_j (t)\dt~, \qquad\qquad\qquad j =1,\dots,n ~,\cr
\d p_1 (t) \,&=\, -\nabla_{q_1}V(q(t))\dt  +\smm r_\Lm(t)\dt~,\cr
\d p_j (t) \,&=\, -\nabla_{q_j }V(q(t))\dt ~,\qquad j =2,\dots,n-1 {}~,\cr
\d p_n (t) \,&=\, -\nabla_{q_n}V(q(t))\dt  +\smm  r_\Rm (t)\dt ~,\cr
\d r_\Lm (t) \,&=\, -\gamma_\Lm  r_\Lm(t)\dt + \lambda_\Lm^2
\gamma_\Lm \,\,q_1(t) \dt
                 - \lambda_\Lm\sqrt{{2\gamma_\Lm\,}/{\beta_\L}}\, \,\d w_\Lm(t)  ~,\cr
\d r_\Rm (t) \,&=\, -\gamma_\Rm  r_\Rm (t)\dt + \lambda_\Rm^2
\gamma_\Rm q_n(t) \dt
                 - \lambda_\Rm \sqrt{{2\gamma_\Rm}/{\beta_\R}}\, \d w_\Rm(t)~,  \cr
}\EQ{eqmo3}
$$
which defines a Markov diffusion process on $\real^{d(2n+2)}$.
\CLAIM{Theorem}{1}{{\rm
[\cite{EPR1999a},\cite{EPR1999b},\cite{RT2000},\cite{EH2000}]} \it
There is a constant $\lambda^* > 0$, such that when
$|\lambda_\Lm|$, $|\lambda_\Rm| \in(0, \lambda^*)$,
the solution of
\equ{eqmo3} is a Markov process
which has an absolutely continuous invariant measure
$\mu$ with a $\CC^{\infty}$ density $m$. This measure is {\bf unique},
{\bf mixing} and attracts any other measure at an exponential rate.}

\REMARKNR{3}$\,$ On can show even a little more. Let $h_0(\beta )$ be the Gibbs distribution for the case where
both reservoirs are at temperature $1/\beta $. If $h$ denotes the density of the invariant measure found in
\clm{1}, we find that $h/h_0(\beta ) $ is in the Schwartz space $\SS$ for all $\beta<\min(\beta_\L,\beta_\R)$.
This mathematical statement reflects the intuitively obvious fact that the chain can not get hotter than either of
the reservoirs.

\REMARKNR{4}$\,$ The restriction on the couplings $\lambda_\Lm$, $\lambda _\Rm$ between the small system and the
reservoirs is a condition of stability (against ``explosion'') of the small system coupled to the heat reservoirs:
It is {\em not} of perturbative nature. Indeed, the reservoirs have the effect of renormalizing the deterministic
potential seen by the small system and this potential must be stable. This restricts $\lambda _\L$ and $\lambda
_\R$.

The proof of \clm{1} is based on a detailed study of
Eq.\equ{eqmo3}. Let $x=(p,q,r)$ and $r=(r_\L,r_\R)$.
For a Markov process $x(t)$ with phase space $X$
and an invariant measure $\mu({\dx})$, ergodic
properties may be deduced from the study of the associated semi-group $T^t$
on the Hilbert space $\L ^2(X,\mu({\dx}))$. To prove the existence
of the invariant measure in \clm{1} one proceeds as follows:
Consider first
the semi-group $T^t$ on the auxiliary Hilbert space $\HH_0\equiv\L
^2(X,\mu_0({\dx}))$,
where the reference measure $\mu_0({\dx})$ is a generalized Gibbs
state for a suitably chosen reference temperature. Our main technical result
consists in proving that the generator $L$
of the semi-group $T^t$ on $\HH_0$ and its
adjoint have compact resolvent.
This is proved by generalizing H\"ormander's techniques for hypoelliptic
operators of ``Kolmogorov type'' to the problem in unbounded domains
described by
\equ{eqmo3}.
Once this is established, we deduce the existence
of a solution to the eigenvalue equation $(T^t)^{*}g=g$ in
$\HH_0$ and this implies immediately the existence of
an invariant measure. The original proof [\cite{EPR1999a}] was subsequently
improved by using more probabilistic techniques [\cite{RT2000}].

The proof of uniqueness, [\cite{EPR1999b}], relies on global
controllability of \equ{eqmo3}. In it, one shows that the control
equation, in which the noises $w_i$ of \equ{eqmo3} are replaced by
deterministic forces $f_i$ (in the same function space), allows
one to reach any given point in phase space in any prescribed
time, by choosing the forces $f_i$ adequately. It is here that, at
least at the time of this writing, a feature of the problem seems
crucial for success:

\REMARKNR{44}$\,$ The geometry of the chain: If the chain is not of linear geometry, but with  parallel strands,
or if the coupling is not of pure nearest neighbor type, uniqueness of the invariant measure does in general not
follow from the methods described here. Very simple counterexamples with harmonic chains [\cite{Zabey2001}] show
that this problem is not easy.

\REMARKNR{analyticity}$\,$ We proved in [\cite{EPR1999a}, Lemma 3.7] that the density $\rho=\rho_T$ is a real
analytic function of $\zeta=(T_\L-T_\R)/(T_\L+T_\R)$. In particular, this yields the standard perturbative results
near equilibrium ($\zeta=0$).

\LIKEREMARK{Question}A fascinating problem is to understand the
limit of a chain of infinitely many oscillators, and in particular
the nature of heat conduction in this case. I believe that this
problem can only be solved if a better understanding of  modeling
the coupling between the heat bath and the chain can be found.

\head{2. Time-reversal, energy flux, and entropy production} \SECTIONcount=2 \EQNcount=1

In the wake of the seminal work of Gallavotti and Cohen [\cite{GC1995}], several
authors realized ({\sl e.g.}, [\cite{K1998}, \cite{LS1999}]) that internal symmetries of
stationary non-equilibrium problems lead to an interesting relation
for the fluctuations in the stationary state. The model we
consider here is no exception, and it is one of the few examples where
the Hamiltonian dynamics plays a very nice role.

It will be useful to streamline the notation.\footnote{${}^1$}{This
notation generalizes more easily to the case $M>1$ of \equ{MM} .}
The two reservoirs, $\L$ and $\R$, are
described by the variables $r=(r_\L,r_\R)\in \real^{d}\oplus\real^{d}$.
Let $\Lambda$ be the ($2d\times nd$) matrix defined by
$$
q\cdot \Lambda r\,=\,
q_1\cdot \Lambda_\L r_\L + q_n \cdot \Lambda_\R r_\R
\,=\,
q_1\smm \lambda _\Lm r_\Lm
+q_n \smm \lambda _\Rm r_\Rm~.
$$
Define the ($2d\times 2d)$ matrix $\Gamma={\rm
diag\ }\gamma_\Lm\oplus{\rm diag\ }\gamma_\Rm$, let
$w=w_\L\oplus
w_\R$  the
$2d$-dimensional standard Brownian motion, and finally $T$ the
($2\times2$) diagonal temperature matrix
$
T\,=\,{\rm diag}(T_\L,T_\R).
$
It is useful to introduce the change of variables
$s=Fr - F^{\T} q$,
where $F= \Lambda \Gamma^{-1/2}$. In terms of
these variables, one can introduce the effective potential
$$
V_\eff(q)\,=\,V(q) -\HALF q\cdot \Lambda \Lambda ^\T q~,
\EQ{Veff}
$$
and the ``energy'' is now $G(s,q,p)$  with
$$
G(s,q,p)\,=\, \HALF p^2 +V_\eff+ \HALF s\cdot \Gamma s~.
\EQ{gdef}
$$
Finally, with the adjoint change in the derivatives $\nabla_q \to
\nabla_ q - F\nabla _s$, the equations of motion \equ{eqmo3} read
$$
\eqalign{
\d q\,&=\,\nabla_p G \dt    \,=\,  p\, \d t~,\cr
\d p\,&=\,-(\nabla_q- F\nabla_s) G \dt \,=\, -\bigl (\nabla_q
V_\eff(q) - F \Gamma s
\bigr )\dt~,\cr
\d s\,&=\, -(\nabla_s +F^\T\nabla_p) G\,\dt - (2T^{1/2}) \d w \,=\,
-\bigl (\Gamma s+ F^\T p\bigr )\dt  - (2T^{1/2}) \d w~. \cr
}
\EQ{final}
$$
Writing $G_p$ for $\nabla_p
G$ and $G_q$ for $\nabla_q G$ (these are vectors with $nd$
components),
and
$G_s$ for $\nabla_s G$ (this is a vector with $2d$ components),
the generator $L$ of the
diffusion process takes, in the variables $y=(s,q,p)$, the form
$$
L\,=\,\nabla_s\cdot T\nabla_s - G_s \cdot \nabla_s + \bigl (
G_p\cdot \nabla_q-G_q\cdot\nabla_p
\bigr )
+\bigl ((FG_s) \cdot \nabla_p-G_p\cdot F\nabla_s\bigr )~.
\EQ{ldef}
$$
If $f$ is a function on the phase space $X$, we let
$$
S^t f(y)\,\equiv\,\bigl (e^{Lt} f\bigr )(y)\,=\,
\int f\bigl (\xi_y(t)\bigr ) d{\bf P}(w)~.
$$
The adjoint $L^\T$ of $L$ in
the space $\L^2(\real^{d(2+2n)})$ is called the Fokker-Planck operator.
The density $ m $ of the invariant measure is the (unique)
normalized
solution of
the equations
$
L^\T  m \,=\,0.
$

\font\boldgreek = cmmib10 scaled 1100

\SUBSECTION{The entropy production {\boldgreek \char27}}

Using the notation \equ{ldef}, we now establish a relation between the
energy flux and the entropy production. Since we are dealing with a
Hamiltonian setup, the energy flux is defined naturally by the time
derivative of the mean evolution $S^t$ of the
effective energy, $H_\eff(q,p)= p^2/2+V_\eff(q)$.
Differentiating, we get from the equations of motion
$
\partial_t  S^t H_\eff\,=\, S^t L  H_\eff~,
$ with
$$
LH_\eff\,=\,  p\cdot (-\nabla_q V_\eff +  F\Gamma s)
+\nabla_q V_\eff\cdot p \,=\,p\cdot F\Gamma s~.
$$
We define the total flux by $\Phi=p\cdot F\Gamma s$, and inspection of
the definition of $F$ and $\Gamma $ leads to the identification of the
flux at the left and right ends of the chain:
$
\Phi\,=\,\Phi_\L+\Phi_\R~,
$
with
$$
\Phi_\L\,=\,p_1\cdot \Lambda _\L \Gamma _\L^{1/2} s_\L~,\qquad
\Phi_\R\,=\,p_n\cdot \Lambda _\R \Gamma _\R^{1/2} s_\R~.
$$
Note that $\Phi_\L$ is the energy flux from the left bath to the
chain,
and $\Phi_\R$ is the energy flux from the right bath to the chain.
Furthermore, observe that
$
\langle \Phi \rangle _\mu\,=\,0~,
$
with
$
\langle f\rangle_\mu\,\equiv\,\int \mu(\d y)\, f(y)=\int \d y\, m(y) f(y)~,
$
because
$\Phi=L H_\eff$ and $L^\T m=0$.

Since we have been able to identify the energy flux on the ends
of the chain, we can {\em define} the (thermodynamic) entropy production
$\sigma$ by
$$
\sigma\,=\, {\Phi_\L\over T_\L}+ {\Phi_\R\over T_\R}\,=\, p\cdot F
T^{-1}\Gamma s~.
\EQ{entropy}
$$

\SUBSECTION{Time-reversal, generalized detailed balance condition}

We next define the ``time-reversal'' map $J$ by
$\bigl (Jf\bigr )(s,q,p)=f(s,q,-p)$. This map is the
projection onto the space of the $s,q,p$ of the
time-reversal of the
Hamiltonian flow (on the full phase space of chain plus baths)
defined by the original problem \equ{eqmo1}.

\LIKEREMARK{Notation}To obtain simple formulas for the entropy
production $\sigma$
we write the (strictly positive) density $ m $ of the invariant
measure $\mu$ as
$$
 m \,=\,Je^{-R}e^{-\varphi}~,
\EQ{phidef}
$$
where
$
R\,=\,R(s)\,=\,\HALF s\cdot \Gamma T^{-1}s~.
$
Let $L^*$ denote the adjoint of $L$ in the space $\HH_\mu=\L^2(X,\d
\mu)$ associated with the
invariant measure $\mu$, where
$X=\real^{2(2n+2)}$.
In terms of the
adjoint $L^\T$ on $\L^2(X,\dspq)$, we have the operator identity
$$
L^*\,=\, m ^{-1} L^\T  m ~.
\EQ{lstar}
$$
We have the following important symmetry property as suggested by
the paper [\cite{K1998}].
\CLAIM{Theorem}{LS1999}\it Let $L_\eta =
L +\eta \sigma$, where $\eta \in \real$. One has the operator
identity
$$
J e^{-J\varphi} (L_\eta )^* e^{J\varphi} J \,=\, L_{1-\eta }~.
\EQ{LS}
$$
In particular,
$$
J e^{-J\varphi} L^* e^{J\varphi} J -L \,=\,\sigma~.
\EQ{LS0}
$$

\REMARKNR{DBC}$\,$ This relation may be viewed as a generalization to non-equilibrium of the detailed balance
condition (at equilibrium, one has $JL^*J-L=0$).

The paper of Gallavotti and Cohen [\cite{GC1995}]
describes fluctuations of the entropy production.
It is based on numerical experiments by [\cite{ECM1990}] which
were then abstracted to the general context of dynamical systems. In
further work, these ideas have been successfully applied to
thermostatted systems modeling non-equilibrium problems.
In the papers [\cite{K1998}] and  [\cite{LS1999}] these ideas have been further extended
to non-equilibrium models described by stochastic dynamics.
In the context of our model, the setup is as follows:
One considers the observable
$
W(t)\,=\,\int_0^t \d \eta\, \sigma\bigl (\xi_x(\eta )\bigr ) ~.
$
By ergodicity, one finds
$\lim _{t\to\infty } t^{-1}W(t)=\langle \sigma \rangle_\mu$,
for all $x$ and almost all realizations of the Brownian motion
$\xi_x(t)=\xi_x(t,\omega )$.
The rate function $\widehat e$ is
characterized by the relation
$$
\inf _{y\in I} \widehat e(y) \,=\,
-\lim_{t\to\infty } {1\over t} \log {\bf Prob}\left \{ {W(t)\over t
\langle \sigma \rangle_\mu}\in I\right  \}~.
$$
Under suitable conditions it can be expressed as the Legendre
transform of the function
$$
e(\eta )\,\equiv\, -\lim_{t\to\infty } t^{-1}\log \bigl \langle
e^{-\eta W(t)}\bigr \rangle_\mu~.
$$
Formally, $-e(\eta )$ can be represented as the maximal
eigenvalue of $L_\eta $. Observing now
the relation \equ{LS}, one sees immediately that
$$
e(\eta)\,=\,e(1-\eta )~.
\EQ{symm}
$$
\vskip -0.5cm \CLAIM{Theorem}{RBT}{{\rm[\cite{RT2001}]} \it The
above relations can be rigorously justified and lead to
$$
\widehat e(y) - \widehat e(-y)\,=\, -y  \langle \sigma \rangle_\mu~.
\EQ{symm1}
$$
}

\vskip -0.8cm
\noindent\rm In particular this means that at equal
temperatures, when $\langle \sigma \rangle_\mu=0$, the
fluctuations are symmetric around the mean $0$, while at unequal
temperatures, the odd part is linear in $y$ and proportional to
the mean entropy production. Note that when $\langle \sigma
\rangle_\mu\ne 0$ this relation describes fluctuations around 0,
{\em not} around the mean! This is the celebrated Gallavotti-Cohen
fluctuation theorem.

\head{3. Extensions}

The technique for proving uniqueness results presented above can
be generalized and applied to many other problems, in particular
to certain types of ``partially noisy'' PDE's (so that now phase
space is infinite dimensional). One kind of example must suffice
to illustrate the kind of results one can obtain. Consider the
stochastic Ginzburg-Landau equation with periodic boundary
conditions (written in Fourier components, for $L\gg 1$):
$$
d u_k \,=\, (1-(k/L)^2)u_k\, dt - \sum_{k_1+k_2+k_3=k}
u_{k_1}u_{k_2}u_{k_3}\, dt + {q_k} \,d w_k~,\quad
k\in {\bf Z}~,
\EQ{hh1}
$$
with $|q_k|\sim k^{-5}$ and where $w_k$ are standard Wiener processes.
The point is here that $q_k$ may be zero for
all $|k|\le k_*$.

\CLAIM{Theorem}{hh1}{{\rm [\cite{EH2001}, \cite{H2002},
\cite{H:2001}]}\it  The process defined by \equ{hh1} has a unique
invariant measure. Any initial condition is attracted
exponentially fast to it.}\rm

\LIKEREMARK{Acknowledgments}I thank M.~Hairer and L.~Rey-Bellet
for help in preparing this manuscript. This work was supported by
the Fonds National Suisse.

\frenchspacing
\head{References}
\endref
\if F\preliminary

\fi
\end{document}